\documentclass[
final
]{dmtcs-episciences}

\usepackage[utf8]{inputenc}

\usepackage[round]{natbib}

\usepackage{amsmath}
\usepackage{amsfonts}
\usepackage{latexsym}
\usepackage{amsmath}

\usepackage{url}

\newtheorem{theorem}{Theorem}[section]

\newtheorem{corollary}[theorem]{Corollary}

\newtheorem{observation}[theorem]{Observation}
\newtheorem{definition}[theorem]{Definition}
\newtheorem{proposition}[theorem]{Proposition}

\author{Daniel J. Rosenkrantz\affiliationmark{1}
  \and Madhav V. Marathe\affiliationmark{2}
  \and S.~S.~ Ravi\affiliationmark{1}
  \and Richard E. Stearns\affiliationmark{1}
}
\title[Symmetry and NCFs]
      {Symmetry Properties of Nested Canalyzing Functions}
\affiliation{
  Biocomplexity Institute and Initiative, University of Virginia and
  Department of Computer Science, University at Albany 
  -- State University of New York, USA\\
  Biocomplexity Institute and Initiative \& Department of 
  Computer Science, University of Virginia, USA
}
\received{2019-6-27}
\revised{2019-10-2}
\accepted{2019-10-29}

\begin{document}
\publicationdetails{21}{2019}{4}{19}{5565}
\maketitle



\newcommand{\QED}{\hfill\rule{2mm}{2mm}}

\newcommand{\cpoly}{\textbf{P}}
\newcommand{\cnp}{\textbf{NP}}
\newcommand{\cnump}{\textbf{\#P}}
\newcommand{\wtg}{\mbox{$\mathcal{G}$}}

\newcommand{\arr}{\mbox{$\:\longrightarrow\:$}}

\smallskip

\begin{abstract}
Many researchers have studied symmetry properties of 
various Boolean functions. 
A class of Boolean functions, 
called \textbf{nested canalyzing functions} (NCFs),
has been used to model certain biological phenomena.
We identify some interesting relationships between NCFs, symmetric 
Boolean functions and a generalization of symmetric Boolean functions,
which we call $r$-symmetric functions (where $r$ is the symmetry level).
Using a normalized representation for NCFs, we develop a 
characterization of when two variables of an NCF are symmetric.
Using this characterization, we show 
that the symmetry level of an NCF $f$
can be easily computed given a standard representation of $f$.
We also present an algorithm for testing whether 
a given $r$-symmetric function is an NCF.
Further, we show that for any NCF $f$ with $n$ variables, the notion of
strong asymmetry considered in the literature is equivalent to
the property that $f$ is $n$-symmetric. 
We use this result to derive a closed form expression for the
number of $n$-variable Boolean functions 
that are NCFs and strongly asymmetric.
We also identify all the Boolean functions that are NCFs 
and symmetric. 

\end{abstract}

\section{Introduction} 
\label{sec:intro}

\subsection{Canalyzing and Nested Canalyzing Functions}
\label{sse:ncf_def}

Research on the symmetry properties of Boolean
functions has received a lot of attention from the
logic design and automated synthesis communities. 
(Additional discussion on this is provided in Section~\ref{sse:related}.)
In this paper, we focus on the symmetry properties of
a class of Boolean functions (called \textbf{nested canalyzing functions})
which are used to model biological phenomena.
To define this class, we start with a simpler
class of functions, called \textbf{canalyzing} Boolean functions.
This class, introduced by \cite{Kauffman-1969}, is defined as follows.

\begin{definition}\label{def:canalyzing}
Given a set $X = $ $\{x_1, x_2, \ldots, x_n\}$ of $n$  Boolean variables,
a Boolean function $f(x_1, x_2, \ldots, x_n)$ over $X$ is \textbf{canalyzing}
if there is a variable $x_i \in X$ and values $a$, $b$ $\in \{0,1\}$ such that
\[
f(x_1, \ldots, x_{i-1}, a, x_{i+1}, \ldots, x_n) ~=~ b, 
\]
for all combinations of values assigned to the variables in $X - \{x_i\}$.
\end{definition}

\medskip

\noindent
\textbf{Example 1:}~ Consider the function 
$f(x_1, x_2, x_3)$ $~=~$ $\overline{x_1} \wedge (x_2 \vee \overline{x_3})$.
This function is canalyzing since if we set to $x_1 = 1$,
$f(1, x_2, x_3) ~=~ 0$, for all combinations of values for 
$x_2$ and $x_3$. 

\medskip

Another class of Boolean functions, called \textbf{nested canalyzing functions} (NCFs),
was introduced later by \cite{Kauffman-etal-2003} to carry out a detailed
analysis of the behavior of certain biological systems.
We follow the presentation in \cite{Layne-2011} in defining NCFs.
(For a Boolean value $b$,~ the complement is denoted by $\overline{b}$.)

\begin{definition}\label{def:nested_canalyzing}
Let $X = $ $\{x_1, x_2, \ldots, x_n\}$ denote a set of $n$  Boolean variables.
Let $\pi$ be a permutation of $\{1, 2, \ldots, n\}$.
A Boolean function $f(x_1, x_2, \ldots, x_n)$ over $X$ is \textbf{nested canalyzing}
in the variable order $x_{\pi(1)}, x_{\pi(2)}, \ldots, x_{\pi(n)}$ with
\textbf{canalyzing values} $a_1, a_2, \ldots, a_n$ and 
\textbf{canalyzed values} $b_1, b_2, \ldots, b_n$ 
if $f$ can be expressed in the following form:
\[
f(x_1, x_2, \ldots, x_n) ~=~ 
   \begin{cases}
       \:b_1 & \mathrm{if~~} x_{\pi(1)} ~=~ a_1 \\
       \:b_2 & \mathrm{if~~} x_{\pi(1)} ~\neq~ a_1 \mathrm{~~and~~}\\
             & x_{\pi(2)} ~=~ a_2 \\
       \:\vdots & \vdots \\ 
       \:b_n & \mathrm{if~~} x_{\pi(1)} ~\neq~ a_1 \mathrm{~~and~~} \ldots \\
             & x_{\pi(n-1)} ~\neq~ a_{n-1} \mathrm{~~and~~}\\
             & x_{\pi(n)} ~=~ a_n \\ 
       \:\overline{b_n} & \mathrm{if~~} x_{\pi(1)} ~\neq~ a_1 \mathrm{~~and~~} \ldots\\
                        & x_{\pi(n)} ~\neq~ a_n. \\
   \end{cases}
\]
\end{definition}
For convenience, we will use a notation introduced in \cite{Stearns-etal-2018}
to represent NCFs.
For $1 \leq i \leq n$, line $i$ of this representation has the form

\medskip

\noindent
\hspace*{0.5in} $x_{\pi(i)}:~ a_i ~~\longrightarrow~~ b_i$

\medskip

\noindent 
where $x_{\pi(i)}$ is the \textbf{canalyzing variable} that is
\textbf{tested} in line $i$, 
and $a_i$ and $b_i$ are respectively the \emph{canalyzing} and 
\emph{canalyzed} values in line $i$,~ $1 \leq i \leq n$.
Each such line is called a \textbf{rule}.
When none of the conditions ``$x_{\pi(i)} ~=~ a_i$" 
is satisfied, we have line $n+1$ with the ``Default" rule
for which the canalyzed value is~ $\overline{b_n}$: 

\medskip

\noindent
\hspace*{0.5in} Default:~ $\overline{b_n}$

\medskip
\noindent
As in \cite{Stearns-etal-2018}, we will refer to the above specification
of an NCF as the \textbf{simplified representation} and assume
(without loss of generality) that each NCF is specified in this manner.
The simplified representation provides the following 
computational view of an NCF. 
The lines that define an NCF are 
considered sequentially in a top-down manner.
The computation stops at the first line where the 
specified condition is satisfied, and the value of the function
is the canalyzed value on that line. 
We now present an example of an NCF using the two representations
mentioned above.

\medskip
\noindent
\textbf{Example 2:}~ Consider the function 
$f(x_1, x_2, x_3) ~=~ \overline{x_1} \wedge (x_2 \vee \overline{x_3})$
used in Example~1.
This function is nested canalyzing using the identity permutation $\pi$ on $\{1,2,3\}$
with canalyzing values $1,1,0$ and canalyzed values $0, 1, 1$.
We first show how this function can be expressed using the syntax of
Definition~\ref{def:nested_canalyzing}.

\[
f(x_1, x_2, x_3) ~=~ 
   \begin{cases}
       \:0 & \mathrm{if~~} x_{1} ~=~ 1 \\
       \:1 & \mathrm{if~~} x_{1} ~\neq~ 1 \mathrm{~~and~~}
            x_{2} ~=~ 1 \\
       \:1 & \mathrm{if~~} x_{1} ~\neq~ 1 \mathrm{~~and~~}
            x_{2} ~\neq~ 1 \mathrm{~~and~~}\\ 
           &x_{3} = 0 \\
       \:0 & \mathrm{if~~} x_{1} ~\neq~ 1 \mathrm{~~and~~} 
             x_{2} ~\neq~ 1 \mathrm{~~and~~}\\ 
           & x_{3} ~\neq~ 0 \\ 
   \end{cases}
\]

\medskip
\noindent
A simplified representation of the same function is as follows.

\bigskip

\noindent
\begin{tabular}{ll}
\hspace*{0.25in} & $x_1:~$  $1 ~\longrightarrow~ 0$ \\ [0.5ex]
\hspace*{0.25in} & $x_2:~$  $1 ~\longrightarrow~ 1$ \\ [0.5ex]
\hspace*{0.25in} & $x_3:~$  $0 ~\longrightarrow~ 1$ \\ [0.5ex]
\hspace*{0.25in} & Default:~ $0$ \\
\end{tabular}

\noindent

\subsection{Symmetric Boolean Functions}
\label{sse:symmetry}

A pair of variables of a Boolean function $f(x_1, x_2, \ldots, x_n)$ is
said to be \textbf{symmetric} if their values can be interchanged without
affecting the value of the function.
As a simple example, each pair of variables in the function 
$f_2(x_1, x_2, x_3)$ = $x_1 \oplus x_2 \oplus x_3$ is symmetric,
where `$\oplus$' represents the Exclusive-Or operator.
This form of interchange symmetry partitions the set of variables into a set of
\textbf{symmetry groups}, where the members of each group are pairwise symmetric.
A Boolean function $f$ is 
$r$-\textbf{symmetric} if it has at most $r$ symmetry groups.
In this case, the value of $f$
depends only on how many of the variables in each symmetry group have the value 1.
We say that $f$ is \textbf{properly} $r$-\textbf{symmetric} if
it is $r$-symmetric, but not $(r-1)$-symmetric.
For example, the function $f_3(x_1, x_2, x_3) = (x_1 \wedge x_2) \vee\, \overline{x_3}$~
is not 1-symmetric since $f_3(1, 0, 1) \neq f_3(1, 1, 0)$; however, 
it is 2-symmetric with the symmetric groups being $\{x_1, x_2\}$ and $\{x_3\}$.
For a Boolean function $f$, the integer $r$ such that $f$ is properly $r$-symmetric
is referred to as the \textbf{symmetry level} of $f$.
Thus, the symmetry level of $f$ is the smallest integer $r$ such that
$f$ is $r$-symmetric.

In the literature (see e.g., \cite{Crama-Hammer-2011,HT-2016,Toth-etal-1977}),
a 1-symmetric function $f$ is referred to simply
as a \textbf{symmetric} function or 
as a \textbf{totally symmetric function} \citep{Biswas-1970,BS-1968}.
Thus, in any symmetric function, each pair of variables is symmetric.
As a simple example, the function $f_2(x_1, x_2, x_3)$ = 
$x_1 \oplus x_2 \oplus x_3$ (defined above) is symmetric.
If $f$ is a symmetric function, then for any
input $(a_1, a_2, \ldots, a_n)$ to $f$, where $a_i \in \{0,1\}$ for
$1 \leq i \leq n$, and any permutation $\pi$ of $\{1, 2, \ldots, n\}$,
$f(a_1, a_2, \ldots, a_n)$ = $f(a_{\pi(1)}, a_{\pi(2)}, \ldots, a_{\pi(n)})$.
For fixed $r \geq 1$, the class of $r$-symmetric functions has been
studied in the literature on discrete dynamical systems (see e.g., 
\cite{Barrett-etal-2007,Rosenkrantz-etal-2015,MR-2007}).

Other forms of symmetry which can be more general than the permutations 
corresponding to symmetry groups have also been considered (see e.g.,
\cite{Maurer-2015,KS-2000}).
A Boolean function $f(x_1, x_2, \ldots, x_n)$
is \textbf{strongly asymmetric}
if for any permutation $\pi$ of $\{1, 2, \ldots, n\}$
\emph{except} the identity permutation,
there exists an input $(a_1, a_2, \ldots,  a_n)$
to $f$ such that $f(a_1, a_2, \ldots, a_n)$ $\neq$
$f(a_{\pi(1)}, a_{\pi(2)}, \ldots, a_{\pi(n)})$.
In general, there are Boolean functions with $n$ variables that are properly
$n$-symmetric, but not strongly asymmetric \citep{KS-2000}.
However, in the case of NCFs with $n$ variables, we will show 
(see Section~\ref{sec:ncf_and_symmetry}) that the notion of 
strong asymmetry coincides with that 
of being properly $n$-symmetric.


\subsection{Summary of Results}
\label{sse:contrib}

Our focus is on the relationships between 
NCFs and symmetric Boolean functions.
Using a normalized representation for NCFs 
(defined in Section~\ref{sec:prelim}), we develop a
characterization of when two variables of an NCF are symmetric.
Using this characterization, we show
that the symmetry level of an NCF $f$
can be easily computed given a normalized representation of $f$.
(In contrast, we show that 
one cannot even efficiently approximate the symmetry level of
a general Boolean function to within any factor $\geq 1$, 
unless \textbf{P} = \cnp.)
We also present an algorithm to test whether a given
$r$-symmetric function $f$ is an NCF. 
Further, we show that for any NCF $f$ with $n$ variables, the notion of
strong asymmetry considered in the literature is equivalent to
the property that $f$ is $n$-symmetric.
We use this result to derive a closed form expression for the
number of $n$-variable Boolean functions
that are NCFs and strongly asymmetric.
In addition, we identify all the Boolean functions that are 
canalyzing and symmetric as well as those that 
are NCFs and symmetric.

\subsection{Related Work}
\label{sse:related}

The usefulness of symmetry properties in synthesizing 
combinational logic functions was first noted 
by \citet{Shannon-1938}.
Over the years, a considerable amount of research on detecting 
and exploiting symmetry properties 
of Boolean functions has been reported in the literature.
For example, \citet{Biswas-1970} and \citet{TM-1996} present methods 
to determine whether a given Boolean function is symmetric.
\citet*{BS-1968} show how certain Boolean functions
can be converted into symmetric functions by introducing additional
variables. 
Several researchers have developed techniques for exploiting symmetries
for automated logic synthesis 
(e.g., \cite{AP-2008,Hu-etal-2008,KS-2000,Darga-etal-2008}).
\citet{Maurer-2015} provides a thorough discussion of known
symmetry detection algorithms and presents a new universal algorithm 
for detecting any form of permutation-based symmetry in Boolean functions. 

The term \textbf{canalization}, coined by
\citet{Waddington-1942}, is generally used to describe
the stability of a biological system with changes
in external conditions.
\citet{Kauffman-1969} introduced canalyzing Boolean functions
to explain the stability of gene regulatory networks.
The subclass of NCFs was
introduced later by \citet{Kauffman-etal-2003} 
to facilitate a rigorous analysis of the Boolean network model
for gene regulatory networks.
It is known that the class of NCFs coincides with that
of unate cascade Boolean functions \citep{Jarrah-etal-2007}.
In the literature on computational learning theory,
NCFs are referred to as $1$-\textbf{decision lists} \citep{KV-1994}.
Many researchers have pointed out the usefulness of NCFs 
in modeling biological phenomena 
(e.g., \cite{Layne-2011,
Layne-etal-2012,Li-etal-2011,Li-etal-2012,Li-etal-2013}).
Properties of NCFs such as sensitivity and 
stability have also been studied in the 
literature (e.g., \cite{Kauffman-etal-2004, Layne-2011,Layne-etal-2012,
Li-etal-2011,Li-etal-2013,Klotz-etal-2013, Stearns-etal-2018,
Paul-etal-2019,2017b-Kadelka-etal}).
Generalized versions of nested canalyzing functions where the 
variables and function values can be from a domain 
of size three or more have also been studied 
(e.g., \cite{2017a-Kadelka-etal}).

To our knowledge, relationships between NCFs and symmetric
Boolean functions have not been addressed in the literature.

\section{Other Definitions and Preliminary Results}
\label{sec:prelim}

\subsection{Overview}

We present some definitions and recall
a known result regarding NCFs which allow us
to define a normalized representation for NCFs.
This representation plays an important role in several
problems considered in later sections.
We also show (Theorem~\ref{thm:approx_sym_level_hard}) 
that, in general, the symmetry level
of a Boolean function cannot be efficiently approximated to within any
factor $\rho \geq 1$, unless \textbf{P} = \cnp.

\medskip

\subsection{Layers and Normalized Representation}
\label{sse:ncf_layer}

Recall that any NCF $f$ with $n$ variables,
denoted by $x_1$, $x_2$, $\ldots$, $x_n$,~
is specified using a variable ordering 
$x_{\pi(1)}$, $x_{\pi(2)}$, $\ldots$,  $x_{\pi(n)}$,
where $\pi$ is a permutation of $\{1, 2, \ldots, n\}$.
Throughout this paper, we will assume without loss of generality
that the variable ordering uses the identity permutation 
so that $x_i$ is the variable tested in line $i$, $1 \leq i \leq n$. 
We assume that an NCF $f$ with $n$
variables is specified using the simplified representation with $n$ lines. 
Unless otherwise mentioned, the ``Default" line (which is assigned the number $n+1$)
in the simplified representation of an NCF is \emph{not} considered
in the results stated in this section.
We will use the following result from \cite{Stearns-etal-2018}.

\begin{observation} \label{obs:ncf_transformations}
Let $f$ be an NCF with $n$ variables specified using $n$ lines
in the simplified representation. 
\begin{enumerate}
\item 
For any $q \geq 2$ and for any $i$, $1 \leq i \leq n-q+1$,
if the $q$ consecutive lines $i$, $i+1$, $\ldots$, $i+q-1$~
have the same canalyzed value, then the function remains
unchanged if these $q$ lines are permuted in any order
without changing the other lines.

\item 
Suppose lines $n-1$ and $n$ in the specification of $f$ 
have complementary canalyzed values.
Then, the function remains unchanged 
if the canalyzing value and canalyzed value in line $n$
are both complemented. 
(Here, the value on the ``Default" line is also complemented.)
\QED
\end{enumerate}
\end{observation}

\citet{Li-etal-2013} defined the concept of a 
{\bf layer} of an NCF in terms of
an algebraic representation of the NCF as an extended monomial.
For our purposes, we use the following definition based on the
simplified representation of NCFs.

\begin{definition}
\label{def:layer}
A {\bf layer} of an NCF representation is a maximal length sequence 
of lines with the same canalyzed value.
\end{definition}

\noindent
\textbf{Example 4:}~ Consider the following function $f$ of six variables
$x_1$, $x_2$, $\ldots$, $x_6$. 

\bigskip

\noindent
\begin{tabular}{ll}
\hspace*{0.25in} & $x_1:~$  $1 ~\longrightarrow~ 0$ \\ [0.5ex]
\hspace*{0.25in} & $x_2:~$  $0 ~\longrightarrow~ 0$ \\ [0.5ex]
\hspace*{0.25in} & $x_3:~$  $0 ~\longrightarrow~ 0$ \\ [0.5ex]
\hspace*{0.25in} & $x_4:~$  $1 ~\longrightarrow~ 1$ \\ [0.5ex]
\hspace*{0.25in} & $x_5:~$  $1 ~\longrightarrow~ 1$ \\ [0.5ex]
\hspace*{0.25in} & $x_6:~$  $1 ~\longrightarrow~ 0$ \\ [0.5ex]
\hspace*{0.25in} & Default:~ $1$ \\
\end{tabular}

\medskip
\noindent
This function has 3 layers, with 
Layer~1 consisting of the lines with $x_1$, $x_2$ and $x_3$, 
Layer~2 consisting of the lines with $x_4$ and $x_5$, and
Layer~3 consisting of the line with $x_6$.  

\medskip
From Part~1 of Observation~\ref{obs:ncf_transformations},
it follows that the lines within the same layer of an NCF
representation can be permuted without changing the function.
Part~2 of Observation~\ref{obs:ncf_transformations} points
out that for any NCF $f$ with $n \geq 2$ variables, there is a simplified 
representation in which lines $n-1$ and $n$ are in the same layer.
We can now define the notion of a \textbf{default-normalized} 
representation for NCFs.

\begin{definition}
\label{def:normalized}
Let $f$ be an NCF with $n \geq 2$ variables specified using
the simplified representation with 
the variable ordering $\langle x_1, x_2, \ldots, x_n\rangle$.
We say that the representation is {\bf default-normalized} if
lines $n-1$ and $n$ have the same canalyzed value.
\end{definition}

\noindent
\textbf{Example 5:}~ 
Consider the NCF $f$ shown in Example~4. 
This representation is \emph{not} default-normalized since lines 5 and 6
have different canalyzed values.
If we change line 6 to ``$x_6$:~ 0 $\longrightarrow$ 1" and the
value on the ``Default" line to 0, then we obtain 
a default-normalized representation. 

\medskip
From Observation \ref{obs:ncf_transformations},
we note that in polynomial time, for any NCF representation with more than one line,
we can first modify the last line if necessary 
so that it has the same canalyzed value as the preceding line,
thereby obtaining an equivalent default-normalized NCF representation.
We state this formally below.

\begin{observation}\label{obs:normalization_poly}
Given an NCF representation for a Boolean function $f$, a default-normalized
representation for $f$ can be obtained in polynomial time. \QED
\end{observation}

\noindent
In view of Observation~\ref{obs:normalization_poly},
we assume henceforth that any NCF is specified in default-normalized form.

\subsection{Complexity of Approximating the Symmetry Level of a 
General Boolean Function}
\label{sse:symmetry_level_hardness}

Recall that the symmetry level of a Boolean function $f$ is the smallest
integer $r$ such that the following condition holds:
the  inputs to $f$ can be partitioned into $r$ subsets 
(symmetry groups) so that the value of $f$ depends only on 
how many of the inputs in each group have the value 1.
We now show that given a Boolean function $f$
in the form of a Boolean expression, it is \cnp-hard to 
approximate\footnote{An algorithm approximates the symmetry
level of a Boolean function $f$ within the factor $\rho$ if it finds a
partition of the inputs to $f$ into at most $\rho \times r^*$ symmetry
groups, where $r^*$ is the symmetry level of $f$.}
the symmetry level of $f$ to within any factor $\rho \geq 1$.
This result holds even when the function is given as an expression
in conjunctive normal form (CNF), that is, in the form of a conjunction
of clauses where each clause is a disjunction of literals. 

\newcommand{\cala}{\mbox{$\mathcal{A}$}}

\begin{theorem}\label{thm:approx_sym_level_hard}
Unless \textbf{P} = \cnp,
for any $\rho \geq 1$, there is no polynomial time 
algorithm for approximating the 
symmetry level of a Boolean function $f$ 
specified as a Boolean expression to within the factor $\rho$.
This result holds even when $f$ is a CNF expression.
\end{theorem}

\noindent
\textbf{Proof:}~ For some $\rho \geq 1$, suppose there is an efficient algorithm \cala{} 
that approximates the symmetry level of a Boolean function
specified as a Boolean expression to within the factor $\rho$.
Without loss of generality, we can assume that $\rho$ is an integer.
We will show that \cala{} can be used to efficiently solve
the CNF Satisfiability problem (SAT) which is known 
to be \cnp-hard \citep{GJ-1979}.

Let $g$ be a CNF formula representing an instance of SAT.
Let $X$ = $\{x_1, x_2, \ldots, x_n\}$ denote the set of
Boolean variables used in $g$.
We create another CNF formula $f$ as follows.
Let $Y = \{y_1, y_2, \ldots, y_{\rho+1}\}$ and
$Z = \{z_1, z_2, \ldots, z_{\rho+1}\}$ be two 
new sets of variables, with each set containing $\rho+1$ variables.
The expression for $f$, which is a function of $n+2\rho+2$
variables, namely $x_1, \ldots, x_n$, $y_1, \ldots, y_{\rho+1}$,
$z_1, \ldots, z_{\rho+1}$, is the following:
\[
    g(x_1, \ldots x_n) \wedge (y_1 \vee \overline{z_1}) 
                       \wedge (y_2 \vee \overline{z_2}) 
                       \wedge \ldots \wedge
                              (y_{\rho+1} \vee \overline{z_{\rho+1}}).
\]
Since $g$ is a CNF formula, so is $f$.
We have the following claims.

\medskip

\noindent
\textbf{Claim 1:}~ If $g$ is \emph{not} satisfiable, 
the symmetry level of $f$ is 1.

\smallskip

\noindent
\textbf{Proof of Claim 1:}~ If $g$ is not satisfiable, $f$ is also
not satisfiable; that is, for all inputs, the value of $f$ is 0.
Thus, all the variables in $X \cup Y \cup Z$ can be included in one
symmetry group.
The value of $f$ is 0 regardless of how many variables in the
group have the value 1.  \hfill$\Box$

\medskip

\noindent
\textbf{Claim 2:}~ If $g$ \emph{is} satisfiable, 
the symmetry level of $f$ is at least $\rho+1$.

\smallskip

\noindent
\textbf{Proof of Claim 2:}~ Suppose $g$ is satisfiable.
We argue that if $i \neq j$, variables $y_i$ and $y_j$
cannot be in the same symmetry group for the function $f$.
To see this, consider any two variables $y_i$ and $y_j$ with $i \neq j$,
and construct an assignment $\alpha$ to
the variables of $f$ in the following manner.
\begin{enumerate}
\item For the variables in $X$, choose any assignment that
satisfies $g$.
\item Let $y_i = 0$, $z_i = 1$, $y_j = 1$ and $z_j = 0$.
\item For $1 \leq p \leq \rho+1$, $p \neq i$ and $p \neq j$,
let $y_p = 1$ and $z_p$ = 0.
\end{enumerate}
Since this assignment $\alpha$ sets the clause $(y_i \vee \overline{z_i})$ to 0,
the value of $f$ under the assignment $\alpha$ is 0.
Now, consider the assignment $\alpha'$ that is obtained by interchanging
the values of $y_i$ and $y_j$ in $\alpha$ without changing the
values assigned to the other variables. 
It can be seen that under assignment $\alpha'$, the value of $f$ is 1.
In both $\alpha$ and $\alpha'$, the number of 1-valued variables in 
the set $\{y_i, y_j\}$ is 1; however, the two assignments lead to 
different values for $f$.
Therefore, if $g$ is satisfiable, for $i \neq j$, $y_i$ and $y_j$ cannot
be in the same symmetry group for $f$.
Since there are $\rho+1$ variables in $Y$, the number of symmetry groups 
for $f$ must be at least $\rho+1$, and this 
completes our proof of Claim~2. \hfill$\Box$

\medskip
 
We now continue with the proof of Theorem~\ref{thm:approx_sym_level_hard}.
Suppose we execute the approximation algorithm \cala{} 
on the function $f$ defined above.
If $g$ is \emph{not} satisfiable, then from Claim~1, 
the symmetry level of $f$ is 1. 
Since \cala{} is a $\rho$-approximation algorithm,
it should produce at most $\rho$ symmetry groups.
On the other hand, if $g$ is satisfiable, by Claim~2, the symmetry
level of $f$ is at least $\rho+1$; so, \cala{} will produce at least
$\rho+1$ groups.
In other words, $g$ is not satisfiable iff the number
of symmetry groups produced by \cala{} is at most $\rho$.
Since \cala{} runs in polynomial time, we have an efficient 
algorithm for SAT, contradicting the assumption that 
\textbf{P} $\neq$ \cnp.  \QED

\medskip

In contrast to the above result,
we will show in the next section that the symmetry level 
of an NCF can be computed efficiently.

\section{Symmetry of Nested Canalyzing Functions}
\label{sec:ncf_and_symmetry}

\subsection{Overview}
\label{sse:res_overview}
Our first result (Theorem~\ref{thm:ncf_symmetric_variables})
provides a characterization of pairs of variables of an NCF
that are part of the same symmetry group. 
This characterization allows us to give a simple closed form
expression (Theorem~\ref{thm:ncf_r_symmetric}) 
for the symmetry level of an NCF specified using
its default-normalized representation.
We also present an algorithm (Theorem~\ref{thm:rsym_canalyzing})
for the converse problem, that is,
testing whether a given $r$-symmetric function $f$ is an NCF;
if so, our algorithm constructs a default-normalized 
representation for $f$.
Next, we show that for any NCF $f$ with $n$ variables,
the two statements ``$f$ is strongly asymmetric" ~and~
``the symmetry level of $f$ is $n$"~ are equivalent 
(Theorem~\ref{thm:ncf_strong_asymmetry}).
We use this result to derive a closed form expression 
for the number of $n$-variable NCFs that are 
also strongly asymmetric (Theorem~\ref{thm:count_strongly_asymmetric}).
Our final result (Proposition~\ref{pro:ncf_symmetric})
identifies all the functions that are symmetric and
canalyzing as well as those that are symmetric and NCFs.

\subsection{Symmetric Pairs of Variables in an NCF:
A Characterization and\newline its Applications}
\label{sse:ncf_strong_sym}

\begin{theorem}\label{thm:ncf_symmetric_variables}
Two variables of an NCF $f$ are symmetric iff
in the default-normalized representation of $f$,
they occur in the same layer and have the same canalyzing value.
\end{theorem}
\noindent
\textbf{Proof:}~
Suppose that two variables of an NCF $f$ occur in the same layer of a
default-normalized representation, and the lines containing these variables
have the same canalyzing value.  Then in any input assignment, the
values of these two variables can be interchanged, and the evaluation
of the lines in that layer will have the same effect.  Thus, the
variables are symmetric.

For the converse, consider the default-normalized NCF representation of $f$. 
As mentioned earlier, we assume without loss of generality that the
canalyzing variable in line $i$ of $f$ is $x_i$, $1 \leq i \leq n$.
Let  $x_j$ and $x_k$ be two variables that occur in different
layers of $f$, or occur in the same layer, but with different
canalyzing values.  Without loss of generality, 
assume that $j < k$, so that the line for
$x_j$ occurs above the line for $x_k$.
Suppose that line $i$ of $f$, $1 \leq i \leq n$, is 

\smallskip

\hspace*{0.25in}
$x_i : a_i ~\longrightarrow~ b_i$. 

\smallskip

\noindent
Consider the following assignment $\alpha$ = $(c_1, c_2, \ldots,  c_n)$ 
to the variables $x_1$, $x_2$, $\ldots$, $x_n$ of $f$.
\begin{enumerate}
\item For $1 \leq i < j$, set $c_i = \overline{a_i}$. 
\item For $i = j$, set $c_j =a_j$. 
\item For $i = k$, set $c_k =\overline{a_j}$. 
\item For $j < i < k$, and for  $k < i \leq n$, 
if $b_i = b_j$,~ then set $c_i =  \overline{a_i}$,
else set $c_i =  a_i$.
\end{enumerate}
Note that $f(\alpha) = b_j$, since the variables in all lines above
line $j$ have the complement of their canalyzing value, and variable
$x_j$ has its canalyzing value.

\medskip

Now, let $\alpha'$ be the assignment that is obtained from $\alpha$
by interchanging the values of variables $x_j$ and $x_k$, so that
after the interchange, variable $x_j$ has value $\overline{a_j}$ and
variable $x_k$ has value $a_j$.  We will show that $f(\alpha') =
\overline{b_j}$, so variables $x_j$ and $x_k$ are not symmetric.

In line $k$ of $f$, canalyzing value $a_k$ is either the same or
the complement of canalyzing value $a_j$, and canalyzed value $b_k$
is either the same or the complement of canalyzed value $b_j$,
Thus, there are four possible cases for the form of line $k$.  We
now consider each of the four cases.

\medskip

\noindent
{\bf Case 1:} Line $k$ has the form $x_k : a_j ~\longrightarrow~ b_j$.  

\smallskip

Since line $k$ has
the same canalyzing value and canalyzed value as line $j$, line
$k$ must occur in a lower layer than that of line $j$.  Thus, there
is at least one line between line $j$ and line $k$ for which the
canalyzed value is $\overline{b_j}$.  Let line $q$ be the first
such line.  Note that for all $i$ such that $1 \leq i < q$, the
value of variable $x_i$ in assignment $\alpha'$ does not match the
canalyzing value of line $i$, but the value of variable $x_q$ does
match the canalyzing value of line $q$.  Since canalyzed value
$b_q$ equals $\overline{b_j}$, we have that $f(\alpha') =
\overline{b_j}$.

\medskip

\noindent
{\bf Case 2:} Line $k$ has the form  $x_k : a_j ~\longrightarrow~ \overline{b_j}$.  

\smallskip

Let line $q$ be the first line such that $j < q \leq k$ and the canalyzed
value of line $q$ is $\overline{b_j}$.  Note that $q$ might possibly
equal $k$.  For all $i$ such that $1 \leq i < q$, the value of
variable $x_i$ in assignment $\alpha'$ does not match the canalyzing
value of line $i$, but the value of variable $x_q$ does match the
canalyzing value of line $q$.  Since canalyzed value $b_q$ equals
$\overline{b_j}$, we have that $f(\alpha') = \overline{b_j}$.

\medskip

\noindent
{\bf Case 3:} Line $k$ has the form $x_k : \overline{a_j} ~\longrightarrow~ b_j$.  

\smallskip

Suppose there is a line below line $j$ for which the canalyzed value is
$\overline{b_j}$.  Let $q$ be the first such line.  Note that for
all $i$ such that $1 \leq i < q$, the value of variable $x_i$ in
assignment $\alpha'$ does not match the canalyzing value of line
$i$, but the value of variable $x_q$ does match the canalyzing value
of line $q$.  Since canalyzed value $b_q$ equals $\overline{b_j}$,
we have that $f(\alpha') = \overline{b_j}$.

\smallskip

Now suppose there is no line below line $j$ for which the canalyzed
value is $\overline{b_j}$.  Then lines $j$ and $k$ both occur in
the last layer of $f$.  Thus, for every line $i$, $1 \leq i \leq
n$, the value of variable $x_i$ in assignment $\alpha'$ does not
match the canalyzing value of line $i$.  Consequently, $f(\alpha')
= \overline{b_j}$, the complement of the canalyzed value in the
last layer of $f$.

\medskip

\noindent
{\bf Case 4:} Line $k$ has the form  $x_k : \overline{a_j} ~\longrightarrow~ \overline{b_j}$. 

\smallskip

Suppose there is a line, other than line $k$, below line $j$ for
which the canalyzed value is $\overline{b_j}$.  Let $q$ be the
first such line.  Note that for all $i$ such that $1 \leq i < q$,
the value of variable $x_i$ in assignment $\alpha'$ does not match
the canalyzing value of line $i$, but the value of variable $x_q$
does match the canalyzing value of line $q$.  Since canalyzed value
$b_q$ equals $\overline{b_j}$, we have that $f(\alpha') =
\overline{b_j}$.

Now suppose that line $k$ is the only line below line $j$ for which
the canalyzed value is $\overline{b_j}$.  Since line $j$ has
canalyzed value $b_j$, line $k$ is the only line in its layer.
Since $f$ is default-normalized, the last layer of $f$ contains at least
two lines.  Thus, there is a last layer following the layer containing
line $k$.  The canalyzed value of all the lines in this last layer
is $b_j$.  Thus, for every line $i$, $1 \leq i \leq n$, the value
of variable $x_i$ in assignment $\alpha'$ does not match the
canalyzing value of line $i$.  Consequently, $f(\alpha') =
\overline{b_j}$, the complement of the canalyzed value in the last
layer of $f$.  \QED

\begin{theorem}\label{thm:ncf_r_symmetric}
Suppose the default-normalized representation of a given NCF contains 
$r_1$ layers with only one distinct canalyzing value,
and $r_2$ layers with two distinct canalyzing values.
Then the function is properly $(r_1 + 2 r_2)$-symmetric.
\end{theorem}

\noindent
\textbf{Proof:}~
From Theorem~\ref{thm:ncf_symmetric_variables}, 
any pair of variables occurring in the same layer, 
with the same canalyzing value, are symmetric.
Thus the function is at most $(r_1 + 2 r_2)$-symmetric.
Moreover, any pair of variables from different layers, 
or with different canalyzing values
are not symmetric, so there are at least $(r_1 + 2 r_2)$ symmetry groups.
\QED

\begin{corollary}\label{cor:ncf_not_rsymm}
For every $n \geq 2$, there is an $n$-variable NCF that is not $n-1$ symmetric. \QED
\end{corollary}

\begin{corollary}\label{cor:ncf_r_symmetric_layers}
(i) An NCF with a default-normalized nested canalyzing representation consisting
of $q$ layers is $2q$-symmetric, and is not $(q-1)$-symmetric.
(ii) An $r$-symmetric NCF has a default-normalized 
nested canalyzing representation with at most $r$ layers. \QED
\end{corollary}

\subsection{Testing Whether an $r$-Symmetric Function is an NCF}
\label{sse:rsym_to_ncf}

Theorem~\ref{thm:ncf_r_symmetric} shows that given an NCF $f$,
the problem of finding the smallest integer $r$ such that
$f$ is $r$-symmetric can be solved efficiently.
We now consider the converse problem, that is, testing whether a given
$r$-symmetric Boolean function is an NCF.
For this problem, we present an algorithm whose running time
is linear in the size of the representation 
of the $r$-symmetric function $f$.
When the answer is ``yes", the algorithm also constructs a
default-normalized representation 
(defined in Section~\ref{sec:prelim}) of $f$.

Let $f(x_1, x_2, \ldots, x_n)$ be an $r$-symmetric 
Boolean function of $n$ variables.
We assume that the symmetry groups 
of $f$, denoted by $g_1$, $g_2$, $\ldots$, $g_r$,
are given.
Let $m_i = |g_i|$, $1 \leq i \leq r$.
Thus, in group $g_i$, the number
of variables which can take on the value 1 varies from 0
to $m_i$, $1 \leq i \leq r$.
We also assume that $f$ is
given as a table $T$ of the following form:
each row of $T$ specifies an $r$-tuple $(c_1, c_2, \ldots, c_r)$,
where $c_i$ is the number of variables of group $g_i$ which
have the value 1, along with the \{0,1\} value of $f$ for that $r$-tuple.
Thus, $\mu$, the number of rows in $T$, 
is given by  $\mu = \prod_{i=1}^r (m_i+1)$.
Since there are $\mu$ rows and each row is an $r$-tuple,
the size of the table $T$ is $O(r\mu)$.
As will be seen, our algorithm for determining whether $f$ is
an NCF runs in $O(r\mu)$ time.
The algorithm relies on the following observation.

\begin{observation}\label{obs:rsym_canalyzing}
Suppose $f(x_1, x_2, \ldots, x_n)$ is an $r$-symmetric Boolean function.
Consider any variable $x_i$ and suppose
the number of variables in the group
containing $x_i$ is $\nu_i$, $1 \leq i \leq n$.
\begin{enumerate}
\item Function $f$ is canalyzing in variable $x_i$
with canalyzing value 1 iff all the table entries where the number
of 1's in the group containing $x_i$ is nonzero have the same value
of $f$. (This value of $f$ is the canalyzed value when $x_i = 1$.)
\item Function $f$ is canalyzing in variable $x_i$ with
canalyzing value 0 iff all the table entries where the number of
1's in the group containing $x_i$ is less than $\nu_i$ have the same
value of $f$. (This value of $f$ is the canalyzed value when $x_i = 0$.)
\QED
\end{enumerate}
\end{observation}
We now explain how Observation~\ref{obs:rsym_canalyzing} can be
used to develop an iterative algorithm for determining whether 
$f$ is an NCF; if so, the algorithm also constructs a default-normalized 
representation for $f$.
We use the following notation.
At beginning of iteration $j$,~
$r_j$ denotes the number of remaining groups,
$X_j$ denotes a set of variables with
exactly one variable from each remaining group,
$T_j$ denotes the table which provides values for the function and
$\mu_j$ denotes the number of rows of $T_j$.
Initially (i.e., $j = 1$),~ $r_1 = r$ (the number of symmetry groups of $f$),
$X_1$ is constructed by choosing one 
variable (arbitrarily) from each of the $r$ groups,
$T_1 = T$ (the given table $T$ for $f$) and 
$\mu_1 = \mu$ (the number of rows of $T$);
further, the NCF representation for $f$ is empty.
The algorithm carries out iteration $j$ if $r_j \geq 1$; 
in that iteration, the algorithm performs Steps I, II and III
as described below.

\smallskip
\noindent
I. Use Observation~\ref{obs:rsym_canalyzing} to determine whether there is
a canalyzing variable $x_i \in X_j$. 

\smallskip
\noindent
II. If yes, perform the following steps. 
\begin{enumerate}
\item Let $\alpha$ and $\beta$ be the respective canalyzing and 
canalyzed values for $x_i$ found in Step~I. 
\item For each variable $x_p$ in the group containing $x_i$, append the rule~
``$x_p : \alpha ~\longrightarrow~ \beta$"~ to the NCF representation of $f$.
\item Set $X_{j+1}$ = $X_j - \{x_i\}$.  
\item Obtain $T_{j+1}$ by retaining only those rows of $T_j$ where all the
variables in the group containing $x_i$ are set to $\overline{\alpha}$. 
\item Set $r_{j+1} = r_j - 1$.
\item If $r_{j+1} \geq 1$, start the next iteration; 
      otherwise, \textbf{stop}.
\end{enumerate}

\smallskip
\noindent
III. (Here, Step~I didn't find a canalyzing variable.)~ 
Output ``$f$ is not an NCF" and \textbf{stop}.

\medskip
\noindent
The correctness of the algorithm is a direct consequence of
Observation~\ref{obs:rsym_canalyzing}. If the algorithm is successful,
it produces a simplified NCF representation of $f$; this can then be
efficiently converted into the default-normalized representation as
discussed in Section~\ref{sse:ncf_layer}.

To estimate the running time, we note that in iteration $j$, we can
determine whether there is a canalyzing variable in $X_j$
in $O(r_j \mu_j)$ = $O(r \mu_j)$ time using 
Observation~\ref{obs:rsym_canalyzing}.
(This is done by considering each variable in $X_j$.)
Once a canalyzing variable is identified, the other steps in that iteration
can be completed in $O(\mu_j)$ time.
Thus, the time for iteration $j$ is $O(r \mu_j)$.
We claim that in each iteration, the number of rows in the table
is reduced by a factor of at least 2; that is, $\mu_{j+1} \leq \mu_j/2$.
To see why, suppose the canalyzing variable $x_i$ found in iteration $j$
is in group $g_p$ with $m_p$ variables. 
Thus, in $T_j$, there are $m_p+1 \geq 2$ possibilities for the number of 1's 
in group $g_p$. 
As indicated in Substep~4 of Step~II above, $T_{j+1}$ 
contains only those rows of $T_j$ where all the variables 
in $g_p$ have the same value.
In other words, there is only one possibility for the number of
1's in group $g_p$.
Thus, $\mu_{j+1}$, the number of rows in $T_{j+1}$,
is at most $\mu_j/(m_p+1)$ $\leq$ $\mu_j/2$.
It follows by simple induction that $\mu_j \leq \mu/2^{j-1}$.
Thus, the running time over all the $r$ iterations is given by
$O(r [\sum_{j=1}^{r}(\mu/2^{j-1})])$ = $O(r\mu)$, since the
geometric sum is bounded by $2\mu$.
Recall that the size of input table $T$ 
(which specifies the $r$-symmetric function $f$)
is $O(r\mu)$. 
Thus, the running time of the algorithm
is linear in the size of the input.
The following theorem summarizes this result. 

\begin{theorem}\label{thm:rsym_canalyzing}
Suppose an $r$-symmetric function $f$ is given as a table where
each row has an $r$-tuple of the form $(c_1, c_2, \ldots, c_r)$,
with $c_i$ being the number of variables in group $g_i$ that have value 1,
along with the \{0,1\} value of the function for that row.
The problem of testing whether $f$ 
is an NCF can be solved in time that 
is linear in the size of the input. \QED
\end{theorem}

When an $r$-symmetric function $f$ with $n$ variables is
specified by giving the table representation described in the
statement of Theorem~\ref{thm:rsym_canalyzing},
the size of the given table is $O(n^r)$.
When $r$ is \emph{fixed}, the size of 
this representation and the running time 
of the above algorithm are both polynomials functions of $n$. 
When $r$ is not fixed, while the size
of the input and the running time of the algorithm are
not necessarily polynomial functions of $n$, 
the running time remains linear in the input size.

\subsection{Strong Asymmetry and NCFs}
\label{sse:strong_asym_ncf}

The notion of strong asymmetry of Boolean functions was defined 
in Section~\ref{sec:prelim}.
For general Boolean functions, the notions of ``$n$-symmetry" and
``strong asymmetry" are not equivalent. 
We illustrate this by presenting an example of a Boolean function
which is properly $n$-symmetric but not strongly asymmetric.

\medskip

\noindent
\textbf{Example 6:}
Consider the Boolean function $f$ of four variables, 
namely $a$, $b$, $c$ and $d$,
whose truth table is shown in Table~\ref{tab:not_str_ssym_ex}.
Function $f$ is symmetric under the permutation $cdab$.
For convenience, we give the truth table for $f$ in a form
that makes its symmetry under the permutation $cdab$ clear.
So, the function $f$ is not strongly asymmetric.
We now demonstrate that the function is properly $4$-symmetric
by observing that it is not symmetric with respect to any pair of variables.

\medskip

\noindent
\begin{minipage}{0.01\textwidth}
\end{minipage}
\begin{minipage}{0.85\textwidth}
\begin{description}
\item{(i)} $f(0100) \neq f(1000)$, so $a$ and $b$ are not symmetric.
\item{(ii)} $f(0011) \neq f(1001)$, so $a$ and $c$ are not symmetric.
\item{(iii)} $f(0001) \neq f(1000)$, so $a$ and $d$ are not symmetric.
\item{(iv)} $f(0100) \neq f(0010)$, so $b$ and $c$ are not symmetric.
\item{(v)} $f(0011) \neq f(0110)$, so $b$ and $d$ are not symmetric.
\item{(vi)} $f(0001) \neq f(0010)$, so $c$ and $d$ are not symmetric.
\end{description}
\end{minipage}


\begin{table}[tb]
\begin{center}
\begin{tabular}{|l|l|c||l|l|c|}\hline
$a~b$ & $c~d$ & {Value of $f$} & $a~b$ & $c~d$ & {Value of $f$} \\ \hline\hline
0~0 & 0~1 & 0 & 0~1 & 1~1 & 1 \\ \hline
0~1 & 0~0 & 0 & 1~1 & 0~1 & 1  \\ \hline
0~0 & 1~0 & 1 & 1~0 & 1~1 & 1 \\ \hline
1~0 & 0~0 & 1 & 1~1 & 1~0 & 1 \\ \hline
0~0 & 1~1 & 0 & 0~0 & 0~0 & 0 \\ \hline
1~1 & 0~0 & 0 & 0~1 & 0~1 & 0 \\ \hline
0~1 & 1~0 & 1 & 1~0 & 1~0 & 0 \\ \hline
1~0 & 0~1 & 1 & 1~1 & 1~1 & 0 \\ \hline
\end{tabular}
\end{center}
\caption{An example of a Boolean function with 4 variables which is
properly $4$-symmetric but not\newline strongly asymmetric.}
\label{tab:not_str_ssym_ex}
\end{table}


\medskip

\noindent
We now show that any $n$ variable NCF that is properly $n$-symmetric
is also strongly asymmetric.
As a consequence, we observe that there are NCFs with $n$ variables
that are properly $n$-symmetric.

\begin{theorem}\label{thm:ncf_strong_asymmetry}
An NCF with $n$ variables is strongly asymmetric iff
it is properly $n$-symmetric.
\end{theorem}

\noindent
\textbf{Proof:}~ 
If an NCF is $(n-1)$-symmetric, then consider a permutation that
interchanges two variables from a symmetry group with at least two
members.  The value of the function is invariant under any such
permutation, so the function is not strongly asymmetric.

For the converse, consider an $n$ variable NCF that is properly $n$-symmetric.
Let $f$ be the default-normalized NCF representation for the function.  For
purposes of notational simplicity, without loss of generality,
we assume that the canalyzing
variable in line $i$ of $f$ is $x_i$, $1 \leq i \leq n$.  Let $\pi$
be any permutation of $\{1, 2, \ldots, n\}$ \emph{except} the
identity permutation.  We will construct an assignment $(c_1, c_2,
\ldots,  c_n)$ to the variables of $f$ such that $f(c_1, c_2, \ldots,
c_n)$ $\neq$ $f(c_{\pi(1)}, c_{\pi(2)}, \ldots, c_{\pi(n)})$.

Let $k$ be the smallest index such that $k \neq \pi(k)$.  Since
$\pi$ is a permutation, $k < n$.  Overall, for $1 \leq i < k$,
$\pi(i) = i$,~ $\pi(k) > k$,~ and for $k < i \leq n$,~ $\pi(i) \geq k$.

\smallskip

Suppose that line $i$ of $f$, $1 \leq i \leq n$, is 

\smallskip

\hspace*{0.25in} $x_i : a_i ~\longrightarrow~ b_i$. 

\smallskip

\noindent
Assignment $(c_1, c_2, \ldots,  c_n)$ is constructed as follows.
\begin{enumerate}
\item For $1 \leq i < k$, set $c_i = \overline{a_i}$.  
\item For $i = k$, set $c_k =a_k$. 
\item For $i > k$, let $i' = \pi^{-1}(i)$, so that $\pi(i')
= i$.  Thus, after the permutation of values, variable $x_{i'}$
will have value $c_i$.  If $b_{i'} = b_k$, then set $c_i =
\overline{a_{i'}}$, else set $c_i =  a_{i'}$.
\end{enumerate}
Since $c_k$ matches the canalyzing value of line $k$ of $f$, and
$c_i$ does not match the canalyzing value of any other earlier line
$i$, $1 \leq i < k$, we have that $f(c_1, c_2, \ldots,  c_n) = b_k$.

Now consider $f(c_{\pi(1)}, c_{\pi(2)}, \ldots, c_{\pi(n)})$.  We
first note that for $1 \leq i < k$, $c_{\pi(i)}$ does not match the
canalyzing value of line $i$.

Let $k' = \pi^{-1}(k)$, so that $\pi(k') = k$.  Canalyzing value
$a_{k'}$ is either the same or the complement of $a_k$, and canalyzed
value $b_{k'}$ is either the same or the complement of $b_k$, Thus,
there are four possible cases for the form of line $k'$ of $f$.  We
will show that in each of the four cases, $f(c_{\pi(1)}, c_{\pi(2)},
\ldots, c_{\pi(n)}) = \overline{b_k}$.

\medskip

\noindent
{\bf Case 1:} Line $k'$ has the form  $x_{k'} : a_k ~\longrightarrow~ b_k$.  

\smallskip

Since line $k'$
has the same canalyzing value and canalyzed value as line $k$, and
$f$ is properly $n$-symmetric, line $k'$ must occur in a lower layer
than that of line $k$.  Thus, there is at least one line between
line $k$ and line $k'$ for which the canalyzed value is $\overline{b_k}$.
Let line $q$ be the first such line.  Note that for all $i$ such
that $1 \leq i < q$, $c_{\pi(i)}$ does not match the canalyzing
value of line $i$, but $c_{\pi(q)}$ does match the canalyzing value
of line $q$.  Since canalyzed value $b_q$ equals $\overline{b_k}$,
we have that $f(c_{\pi(1)}, c_{\pi(2)}, \ldots, c_{\pi(n)}) =
\overline{b_k}$.

\medskip

\noindent
{\bf Case 2:} Line $k'$ has the form $x_{k'} : a_k ~\longrightarrow~ \overline{b_k}$.

\smallskip

Let line $q$ be the first line such that $k < q \leq k'$ and
the canalyzed value of line $q$ is $\overline{b_k}$.
Note that for all $i$ such that $1 \leq i < q$, 
$c_{\pi(i)}$ does not match the canalyzing value of line $i$,
so $f(c_{\pi(1)}, c_{\pi(2)}, \ldots, c_{\pi(n)}) = \overline{b_k}$.

\medskip

\noindent
{\bf Case 3:} Line $k'$ has the form $x_{k'} : \overline{a_k} ~\longrightarrow~ b_k$. 

\smallskip

Suppose there is a line below line $k$ for which the canalyzed
value is $\overline{b_k}$.  Let $q$ be the first such line.  Note
that for all $i$ such that $1 \leq i < q$, $c_{\pi(i)}$ does not
match the canalyzing value of line $i$, but $c_{\pi(q)}$ does match
the canalyzing value of line $q$.  Thus, $f(c_{\pi(1)}, c_{\pi(2)},
\ldots, c_{\pi(n)}) = \overline{b_k}$.

Now suppose there is no line below line $k$ for which the canalyzed
value is $\overline{b_k}$.  
Then line $k$ occurs in the last layer of $f$.  
Since $f$ is both default-normalized and properly $n$-symmetric,
this last layer contains exactly two lines.  Thus $k = n-1$ and $k'
= n$.  Note that for all $i$, $c_{\pi(i)}$ does not match the
canalyzing value of line $i$.  Thus, $f(c_{\pi(1)}, c_{\pi(2)},
\ldots, c_{\pi(n)}) = \overline{b_k}$.

\medskip

\noindent
{\bf Case 4:} Line $k'$ has the form $x_{k'} : \overline{a_k} ~\longrightarrow~ 
              \overline{b_k}$. 

\smallskip

Suppose there is a line, other than line $k'$, below line $k$ for
which the canalyzed value is $\overline{b_k}$.  Let $q$ be the
first such line.  Note that for all $i$ such that $1 \leq i < q$,
$c_{\pi(i)}$ does not match the canalyzing value of line $i$, but
$c_{\pi(q)}$ does match the canalyzing value of line $q$.  Thus,
$f(c_{\pi(1)}, c_{\pi(2)}, \ldots, c_{\pi(n)}) = \overline{b_k}$.

Now suppose that line $k'$ is the only line below line $k$ for which
the canalyzed value is $\overline{b_k}$.  Since line $k$ has
canalyzed value $b_k$, line $k'$ is the only line in its layer.
Since $f$ is default-normalized, there is a last layer following the layer
containing line $k'$.  Thus, for all $i$, $c_{\pi(i)}$ does not
match the canalyzing value of line $i$; therefore, 
$f(c_{\pi(1)}, c_{\pi(2)}, \ldots, c_{\pi(n)})$ ~=~ $\overline{b_k}$.  \QED

\medskip

\noindent
We now provide
examples of properly $n$-symmetric $n$-variable NCFs satisfying the
conditions of Theorem \ref{thm:ncf_r_symmetric}.
For each $n > 1$, let $f_n$ be the function of $n$ variables, namely
$x_1, \ldots, x_n$, defined by the following formula:

\smallskip

\hspace*{0.5in} $x_1 \vee (\overline{x}_2 \wedge 
    ( x_3 \vee (\overline{x}_4 \wedge (\cdots ))))$

\smallskip

\noindent
For instance,
\begin{align*}
f_6 &~=~  x_1 \vee (\overline{x}_2 \wedge (x_3 \vee (\overline{x}_4 \wedge 
      (x_5 \vee \overline{x}_6))))~~~ \mathrm{and}\\
f_7 &~=~  x_1 \vee (\overline{x}_2 \wedge (x_3 \vee (\overline{x}_4 \wedge (x_5 
          \vee (\overline{x}_6 \wedge  x_7))))).
\end{align*}
Function $f_n(x_1, x_2, \ldots, x_n)$ is the NCF corresponding to
the following NCF representation:

\medskip

\noindent
\hspace*{0.25in}
$x_1:~ 1 ~\longrightarrow~ 1$ \\
\hspace*{0.25in}
$x_2:~ 1 ~\longrightarrow~ 0$ \\
\hspace*{0.25in}
$x_3:~ 1 ~\longrightarrow~ 1$ \\
\hspace*{0.75in}
$\vdots$ 

\smallskip

\noindent
If $n$ is odd, the last line is 

\smallskip

\noindent
\hspace*{0.25in}
$x_n:~ 1 ~\longrightarrow~ 1$ 

\smallskip

\noindent
and if $n$ is even, the last line is 

\smallskip

\noindent
\hspace*{0.25in}
$x_n:~ 1 ~\longrightarrow~ 0$. 

\smallskip

\noindent
Note that if $x_i = 0$ for all $i$, $1 \leq i \leq n$, then the
value of $f_n$ is 0 if $n$ is odd and 1 if $n$ is even.  Also, note
that the above representation is not default-normalized. 
To obtain a default-normalized 
representation, the last line would be changed to have canalyzing
value 0, and the same canalyzed value as the preceding line.

\subsection{Number of Strongly Asymmetric NCFs}
\label{sse:number_strongly_asymmetric}

\begin{theorem}\label{thm:count_strongly_asymmetric}
For any $n \geq 2$, the number of Boolean functions with $n$ variables 
that are both strongly asymmetric and NCFs is~ $n! \, 2^{n-1}$.
\end{theorem}

\noindent
\textbf{Proof:}~
Consider a default-normalized representation of a strongly asymmetric NCF $f$.
The representation has $n-1$ layers.
The first $n-2$ layers each contain one line, and the last layer contains two lines.

Consider the canalyzed values in $f$. The canalyzed value in the
first layer can be either 0 or 1.  The canalyzed value in every
other layer is the complement of the canalyzed value in the preceding
layer.  Thus, there are 2 possibilities for the sequence of canalyzed
values in $f$.

The canalyzing variable in each of the first $n-2$ lines (one per
layer) can be any variable that has not yet occurred in a preceding
line.  The last layer contains the remaining two variables.  Thus,
there are $n! /[n-(n-2)]!$ = $n!/2$ possibilities for the pattern of 
canalyzing variable occurrences in $f$.

For each of the first $n-2$ variables, the canalyzing value for the line
containing that variable can be either 0 or 1.  
For the last two lines, the canalyzing values for the variables 
in those lines must be complements of each other; otherwise, by 
Theorem~\ref{thm:ncf_symmetric_variables}, those two variables 
will be symmetric.
Thus, the number of possible canalyzing values over all the $n$
lines is $2^{n-2} \times 2$ = $2^{n-1}$.

Hence, the number of Boolean functions with $n$ variables that
are both strongly asymmetric and NCFs is equal to 
$2 \times (n!/2) \times 2^{n-1}$ ~=~ $n!\,2^{n-1}$.  \QED

\subsection{Symmetric Canalyzing and Nested Canalyzing Functions}
\label{sse:sym_and_cf_ncf}

\begin{proposition}\label{pro:ncf_symmetric}
(i) The only symmetric canalyzing functions are OR, AND, NOR, NAND, 
the constant function 0 and the constant function 1.
(ii) The only symmetric NCFs are OR, AND, NOR and NAND.
\end{proposition}
\noindent
\textbf{Proof:}~

\smallskip

\noindent
\textbf{Part (i):}~
Suppose that symmetric function $f$ is also a canalyzing function. 
Then there is a variable $x$, and values $a$ and $b$ 
such that whenever $x = a$, function $f$ has value $b$.  
Since $f$ is symmetric, $f$  has
value $b$ whenever any of its variables has value $a$.  
Thus, if at least one of its variables has value $a$, then $f$ has value $b$.  
If $f$ has value $\overline{b}$ when none of its variables has
value $a$, the four possible combinations of values for $a$ and $b$
correspond to the four functions OR, AND, NOR, and NAND.  If $f$
has value $b$ when none of its variables has value $a$, then $f$
is the constant function $b$.

\smallskip

\noindent
\textbf{Part (ii):}~
Suppose that symmetric function $f$ is also an NCF.  
Since constant functions are not NCFs,
$f$ has value $\overline{b}$ when none of its variables has value $a$.
The four possible combinations of values for $a$ and $b$ correspond
to the four functions OR, AND, NOR and NAND, each of which is an
NCF.  \QED


\section{Summary}
\label{sec:summary}

We presented a characterization of when two variables
of an NCF are symmetric and used that characterization
to show that the symmetry level of an NCF can be easily computed.
We also showed that an $n$-variable NCF is strongly asymmetric
iff its symmetry level is $n$.
We presented several corollaries of this result, including
a closed form expression for the number of NCFs which are also
strongly asymmetric.
Thus, our results bring out several interesting relationships 
between NCFs and symmetric Boolean functions.


\acknowledgments
We sincerely thank the reviewer for carefully
reading the manuscript and providing very helpful comments.
This work has been partially supported by
DTRA CNIMS (Contract HDTRA1-11-D-0016-0001),
NSF Grant IIS-1908530, 
NSF Grant OAC-1916805, 
NSF DIBBS Grant ACI-1443054, 
NSF BIG DATA Grant IIS-1633028 and
NSF EAGER Grant CMMI-1745207.
The U.S. Government is authorized to reproduce and
distribute reprints for Governmental purposes notwithstanding
any copyright annotation thereon.

\bigskip

\medskip

\noindent
\textbf{Disclaimer:}~ The views and conclusions contained
herein are those of the authors and should
not be interpreted as necessarily representing the
official policies or endorsements, either expressed
or implied, of the U.S. Government.

\bibliographystyle{abbrvnat}
\bibliography{refs}

\end{document}